\documentstyle[12pt,epsfig]{article}
\def\be{\begin{equation}}
\def\ee{\end{equation}}

\def\lsim{\raise0.3ex\hbox{$<$\kern-0.75em\raise-1.1ex\hbox{$\sim$}}}
\def\gsim{\raise0.3ex\hbox{$>$\kern-0.75em\raise-1.1ex\hbox{$\sim$}}}

\begin{document}

\noindent August 6, 2001 \hfill BI-TP 2001/16

\vskip 1 cm

\centerline{\large{\bf Cluster Percolation and First Order}}

\medskip

\centerline{\large{\bf Phase Transitions in the Potts Model}}

\vskip 1.0cm

\centerline{\bf Santo Fortunato and Helmut Satz}

\bigskip

\centerline{Fakult\"at f\"ur Physik, Universit\"at Bielefeld}
\par
\centerline{D-33501 Bielefeld, Germany}

\vskip 1.0cm

\noindent

\centerline{\bf Abstract:}

\medskip

The q-state Potts model can be formulated in geometric terms, with
Fortuin-Kasteleyn (FK) clusters as fundamental objects. If the phase
transition of the model is second order, it can be equivalently
described as a percolation transition of FK clusters. In this work,
we study the percolation structure when the model undergoes a first
order phase transition. In particular, we investigate numerically the
percolation behaviour along the line of first order phase transitions of the
3d 3-state Potts model in an external field and find that the
percolation strength exhibits a discontinuity along the entire line.
The endpoint is also a percolation point for the FK clusters, but the
corresponding critical exponents are neither in the Ising nor in the
random percolation universality class.

\vskip 1cm

\noindent {\bf 1.\ Introduction}

\bigskip

It is quite well established today that thermal features of physical
systems can in many cases be described through the structural properties
of connected geometric objects, or clusters. This mapping between
thermal aspects of the given model and the geometrical properties of
the clusters turns out to be particularly fruitful for the study of
the critical behaviour of such models. The increase of the correlation
length near the critical point is paralleled by the increase of the
average cluster radius, and its divergence to the formation of an
infinite cluster. Percolation theory \cite{stauff,grimm} is the natural
framework to study the properties of cluster-like structures of a
system.

One of the most basic results in this field \cite{fort} shows that
the $q$-state Potts model (without external field) can be mapped onto a
geometric model: spin configurations become FK cluster configurations by
connecting nearest-neighbouring spins of the same orientation with a
bond probability $p_B=1-\exp(-J/kT)$, where $J$ is the Potts spin-spin
coupling. The mapping is one to one, so that any statement about
thermal properties of the model can be equivalently expressed in terms
of cluster quantities. In particular, the magnetization transition is
equivalent to the percolation transition of FK clusters, a result
originally proved for $q=2$ (Ising model) \cite{conkl} and subsequently
extended \cite{peruggi} to any Potts model which undergoes a second
order phase transition. We recall that the Potts model leads to
continuous transitions only for $q=2,3,4$ in two dimensions and for
$q=2$ also in three and higher dimensions.

The FK transformation is, however, independent of the number $q$ of
different spin states of the model. In particular, it remains valid as
well in those cases in which the model exhibits a first order phase
transition. It is thus natural to ask whether one can establish a
relation between the behaviour of FK clusters and the thermal
properties of the system also for discontinuous phase changes.
In this case, a first order transition persists in the presence
of an external field $H$, as long as $H$ remains smaller than some
critical value $H_c$. So for $0\,{\leq}\,H<H_c$, there is a whole
line of first order phase transitions in the phase diagram of the model.
For $H=H_c$, the transition becomes second order and the critical
exponents are conjectured to belong to the Ising universality class;
for $H>H_c$, the partition function becomes analytic and one has at
most a rapid crossover.

We note that the equivalence of the thermal and the percolation description
provided by the FK transformation is valid only for the case $H=0$. The
presence of an external field can in principle be taken into account by
introducing a "ghost spin" connected to all the normal spins of the system
\cite{swendwa}; the resulting clusters then differ from those defined
in \cite{fort}. Nevertheless, the usual FK clusters show a number of
non-trivial features also in the presence of an external field
\cite{kertesz}-\cite{io}.

Here we want to study by means of Monte Carlo simulations the behaviour
of the FK clusters near the line of first order phase transitions
of the 3-dimensional 3-state Potts model. This model has been
investigated quite extensively, since its phase transition is closely
related to the deconfinement transition of finite temperature QCD
\cite{S&Y}. It exhibits a weak first order phase transition for $H=0$,
which disappears already for quite small values of the external field.
The line of first order phase transitions of this model was recently
studied in detail \cite{sven}, and the position of the endpoint was
determined with great precision. In what follows we shall exploit the
results of this investigation.

\bigskip

\noindent{\bf 2.\ The Fortuin-Kasteleyn Transformation and Percolation
Variables}

\bigskip

The ferromagnetic $q$-state Potts model is defined by the Hamiltonian
\be
{\cal H}\,=\,J\sum_{ij}
(1-\delta_{\sigma_i\sigma_j})
-H\sum_i\delta_{\sigma_i\sigma_h}
\label{potts}
\ee
where $J>0$ is the spin-spin coupling and $H$ the external field; the
$\sigma$'s represent the spin variables of the model and can take on
$q$ different values. The direction of the external field is
specified by the spin variable $\sigma_h$. The partition function
${\cal Z}$ is given by
\be
{\cal Z}(T,H)\,=\,\sum_{\sigma}\exp\Big[-{{\cal H}(\sigma) \over
kT}\Big],
\label{p4}
\ee
with the sum over all spin configurations. By distributing randomly
bonds with probability $p_B$ between all pairs of nearest neighbour
sites in the same spin state, one can rewrite ${\cal Z}(T,H=0)$ in the
form
\be
{\cal Z}=\sum_{n}\Big[\prod_{<ij>,n_{ij}=1}p_B
\prod_{<ij>,n_{ij}=0}(1-p_B)~q^{c(n)}\Big].
\label{p8}
\ee
Here the sum runs over all bond configurations $\{n\}$ ($\{n_{ij}=1\}$:
active bond, $\{n_{ij}=0\}$: no bond), and $c(n)$ is the number of FK
clusters of the configuration. We stress that in Eq.\ (\ref{p8})
the spins of the system do not appear; the partition function is given
entirely in terms of bond configurations.

\par

In this work we are mainly interested in the percolation transition of
the FK clusters. We therefore first recall the relevant variables. The
order parameter is the percolation strength $P$, defined as the
probability that a randomly chosen site belongs to an infinite cluster.
The analogue of the magnetic susceptibility is the average cluster size
$S$,
\be
S\equiv \frac{\sum_{s} {{n_{s}s^2}}}{\sum_{s}{n_{s}s}}~,
\label{clustersize}
\ee
where $n_{s}$ is the number of clusters of size $s$; the sums exclude
the percolating cluster.

\par

For thermal second order phase transitions it is found that near the
critical temperature $T_c$,
\be
P \sim (T_c - T)^{\beta}, ~~~~T\lsim T_c
\label{8};
\ee
\be
S \sim |T-T_c|^{-\gamma}
\label{9}
\ee
where $\beta$ and $\gamma$ are the critical exponents for the
magnetization and the susceptibility, respectively.

\par

To study the percolation transition it is helpful to define also the
percolation cumulant. It is the probability of reaching percolation at
a given temperature and lattice size, i.e., the fraction of
"percolating" configurations. This variable has two remarkable
properties:
\begin{itemize}
\item{if one plots it as a function of $T$, all curves corresponding to
different lattice sizes cross at the same point, which marks the
threshold of the percolation transition;}
\item{the percolation cumulants for different values of the lattice size
$L$ coincide, if considered as functions of
$[(T-T_c)/T_c]L^{1/\nu}$.}
\end{itemize}
These two features allow a rather precise determination of the critical
point already from simulations of the system for two different lattice
sizes. In general, it is preferable to utilize several lattices in
order to evaluate finite size effects and eventual corrections to
scaling. Moreover, by using lattices of very different sizes, the
scaling of the curves leads to a more precise estimate of the critical
exponent $\nu$.

\bigskip

\noindent{\bf 3.\ Results}

\bigskip

We have performed Monte Carlo simulations of the 3d 3-state Potts
model for several lattice sizes ($40^3$, $50^3$, $60^3$, $70^3$)
and for different values of the parameters $\beta=J/kT$ and $h=H/kT$,
using the Wolff cluster update extended to the case of a non-vanishing
external field \cite{nieder}. To identify the clusters of the different
configurations, we used the algorithm of Hoshen and Kopelman
\cite{kopelman}. We have always adopted free boundary conditions and
assumed that a cluster percolates if it connects each pair of
opposite faces of the
lattice. At each iteration, we measured the energy $E$ of the system,
the magnetization $M$, the percolation strength $P$, the average
cluster size $S$ and the size of the largest cluster. We recall that
the magnetization is the fraction of spins pointing in the direction of
the external field. In the case of vanishing field, the majority spin
state of the configuration defines the magnetization: $M$ is given by
the fraction of the sites in such spin state. We have also measured the
size of the largest cluster because it allows us to calculate the
fractal dimension $D$ of the percolating cluster at the critical point.
Although the Wolff algorithm is in general very efficient, we were
forced to perform many updates between consecutive measurements
to reduce appreciably the correlations of the corresponding
configurations. In order to get independent configurations for the
percolation variables, we have taken up to 1000 updates for the
$70^3$ lattice, which made the data production rather slow.

\par

For the analysis near the threshold it turns out to be useful to plot
the time history of $M$ and $P$. For a first order phase transition,
because of the finite size of the lattice, the system tunnels from one
phase to the other, which is the lattice realization of the coexistence
of the two phases of the system. This can allow a visual check of the
order of the percolation transition.

\par

\begin{figure}[htb]
\begin{center}
\epsfig{file=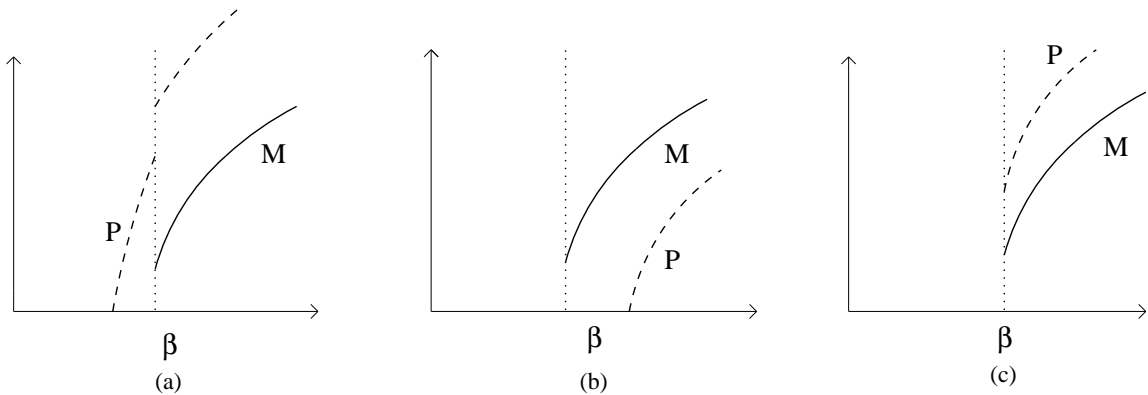,width=15.2cm}
\caption{\label{step}{Scenarios for the relation between percolation
and a first order thermal transition; $M$ denotes the magnetization,
$P$ the percolation strength.}}
\end{center}
\end{figure}

In general, we expect that a step in the magnetization is accompanied
by a step in the average size of the magnetic domains of the system,
and consequently by a step of the FK cluster size as well. It is thus
natural to expect discontinuous variations of the percolation variables
at some threshold. However, the relation between the thermal and the
geometric thresholds is not {\it a priori} evident. There are, in
principle, three possible scenarios:
\begin{itemize}
\item{The configurations of FK clusters percolate at a temperature $T_p$
above $T_c$ ($\beta_p<\beta_c$), and the percolation transition is
continuous with critical exponents; unrelated to this transition,
both $P$ and $S$ then make a jump at $T_c$ (Fig.\ref{step}a).}
\item{The configurations of FK clusters percolate at a temperature $T_p$
below $T_c$ ($\beta_p>\beta_c$), and the percolation transition is
continuous with critical exponents; unrelated to this transition, only
$S$ makes a jump at $T_c$, since $P=0$ there (Fig.\ref{step}b).}
\item{The configurations of FK clusters percolate at $T_c$; the
percolation transition is discontinuous and both $P$ and $S$ make a
jump at $T_c$. In particular, $P$ jumps at $T_c$ from zero to a non-zero
value and is still an order parameter (Fig.\ref{step}c).}
\end{itemize}

For $h=0$, renormalization group arguments suggest that the third
alternative is the correct one, at least in two dimensions ($q>4$)
\cite{peruggi}. What happens in the presence of an external field is so
far not known. Therefore we will present separately our results
for $h=0$ and $0<h<h_c$.

\bigskip

\noindent{\bf 3.1 The Case h=0}

\bigskip

For the model without an external field, the magnetization $M$ is the
order parameter of the thermal transition. By studying the time history
of $M$ at $T_c$, we observe a tunneling between zero and a non-zero
value.

\par

\begin{figure}[htb]
\begin{center}
\epsfig{file=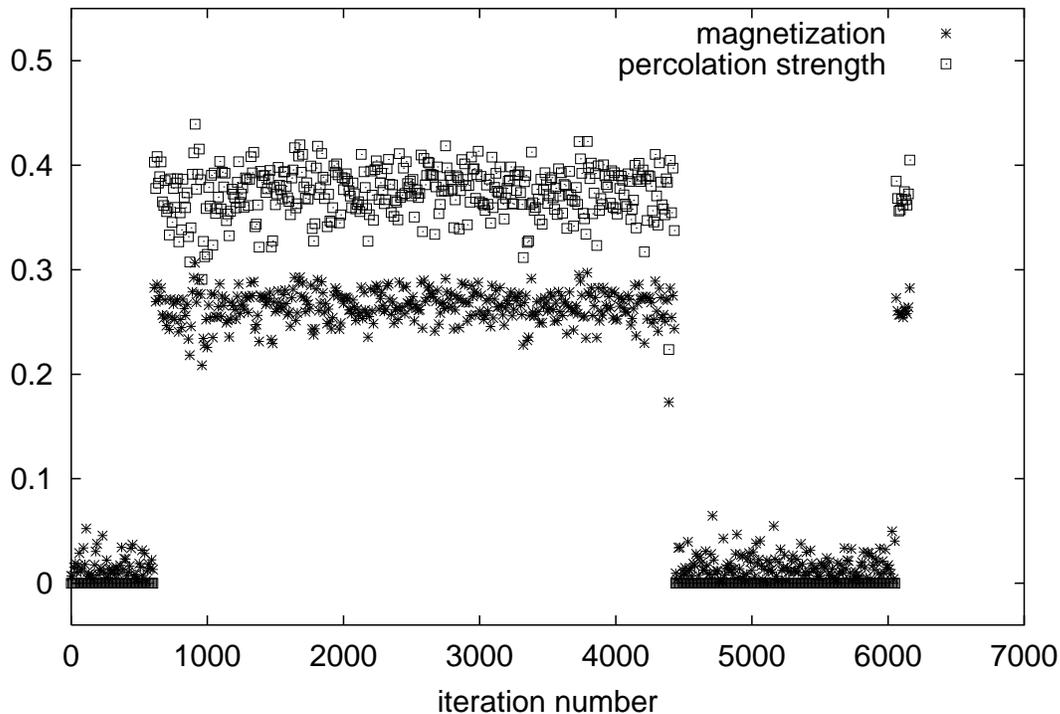,width=16cm}
\caption{\label{mph0}{Time history at the transition point for the 
magnetization $M$ and the percolation strength $P$ of FK clusters 
in the 3d 3-state Potts model without external field. The lattice 
size is $70^3$.}}
\end{center}
\end{figure}

In Fig. \ref{mph0} we show the magnetization $M$ and the percolation
strength $P$ as a function of the number of iteration for a $70^3$
lattice at $\beta_c$. Here we have taken $\beta_c = 0.550565$, as
determined in \cite{janke}. We see that $P$ follows the variations of
$M$; the step of $P$ is somewhat larger and the two "geometric phases"
are clearly visible. In particular, we notice that the system passes
from a non-percolating phase to a percolating one, as conjectured in
\cite{peruggi}. Since the definition of percolation is sharp (either
there is a percolating cluster or there is none), we obtain almost
always $P=0$ when the system is in the non-percolating phase. In
contrast, the magnetization varies smoothly in the lattice average and
hence shows significant fluctuations also in the paramagnetic phase.
Hence $P$ can resolve eventual discontinuities due to different phases
better than $M$ can.

\begin{figure}[htb]
\begin{center}
\epsfig{file=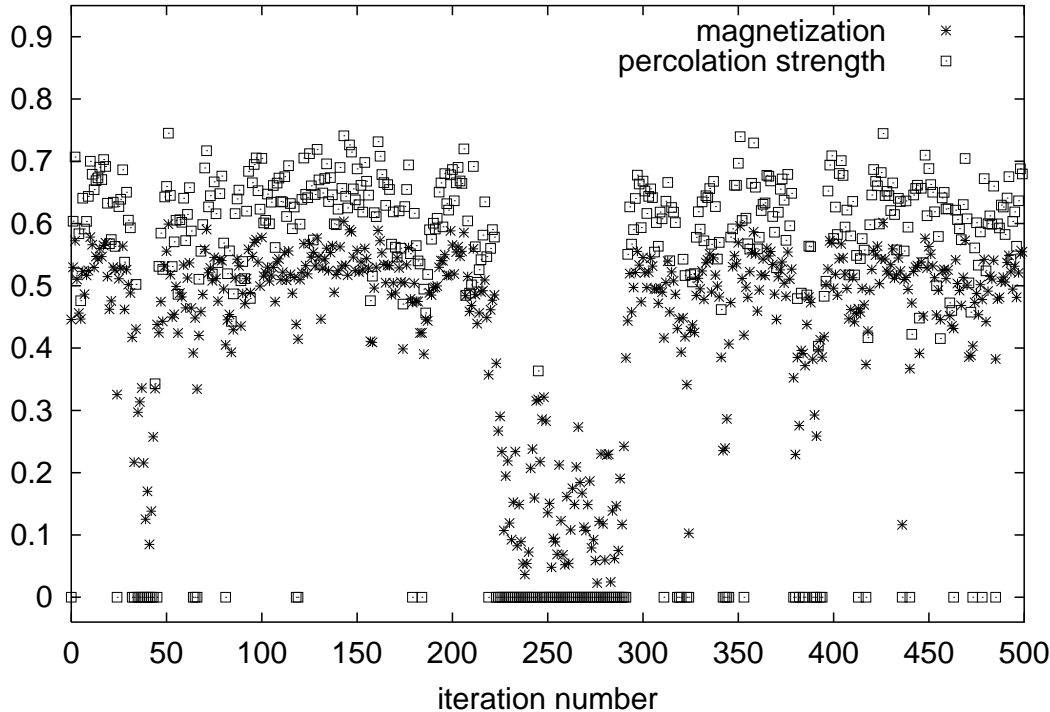,width=16cm}
\caption{\label{mph05}{Time history at the transition point for the
magnetization $M$ and the percolation strength $P$ of FK clusters 
for the 2d 5-state Potts model without external field. The lattice 
size is $200^2$.}}
\end{center}
\end{figure}

To check whether this behaviour of the percolation strength is general,
we repeat our analysis for the 2d 5-state Potts model. In two dimensions
the critical temperature is given by the exact formula
$\beta_c(q)=\log(1+\sqrt{q})$, which for $q=5$ yields
$\beta_c(5)=1.174359$. Fig. \ref{mph05} shows the time history of $M$
and $P$ at $\beta_c(5)$ : the result is the same as before.

\bigskip

\noindent{\bf 3.2 The Case \boldmath{$0<h<h_c$}}

\bigskip

\begin{figure}[p]
\begin{center}
\epsfig{file=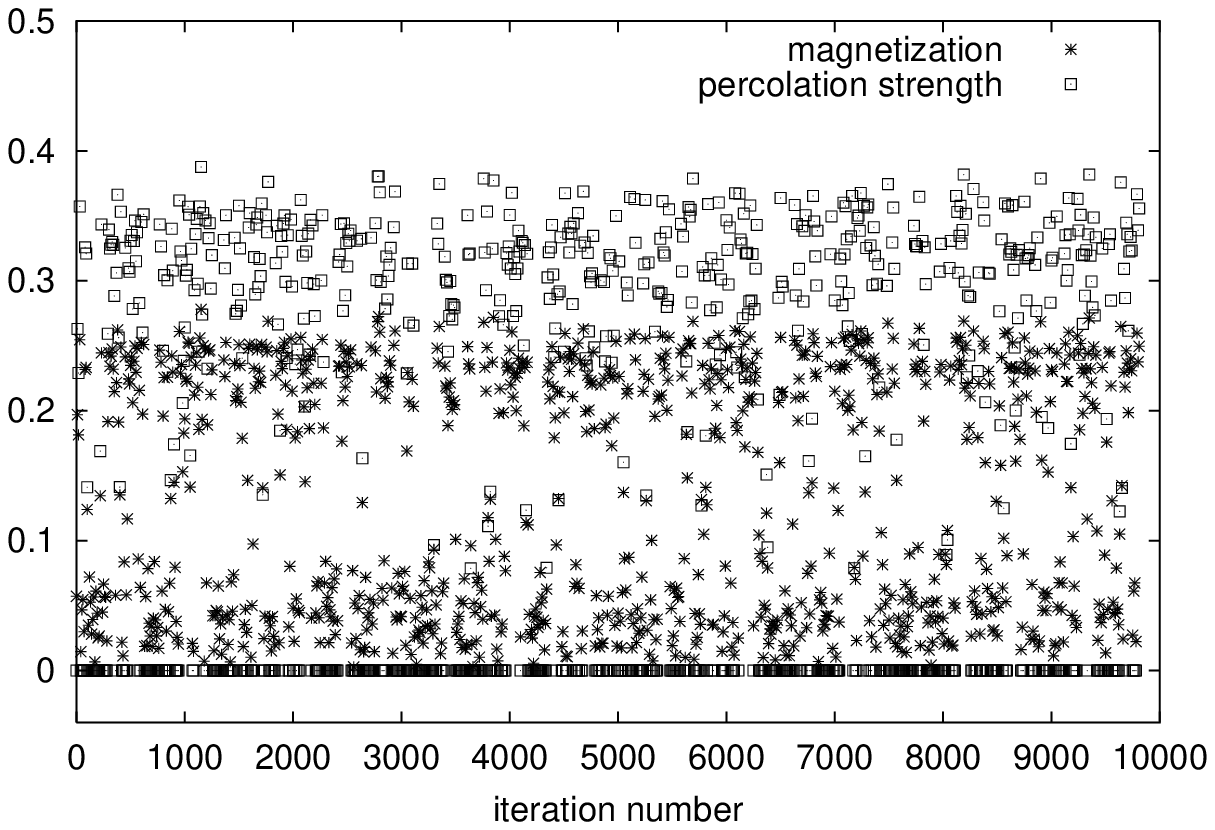,width=12.5cm}
\caption{\label{h1}{Time history at the transition point and with
field $h=0.0005$ for the magnetization $M$ and the percolation 
strength $P$ of FK clusters. The lattice size is $70^3$.}}
\vskip2cm
\epsfig{file=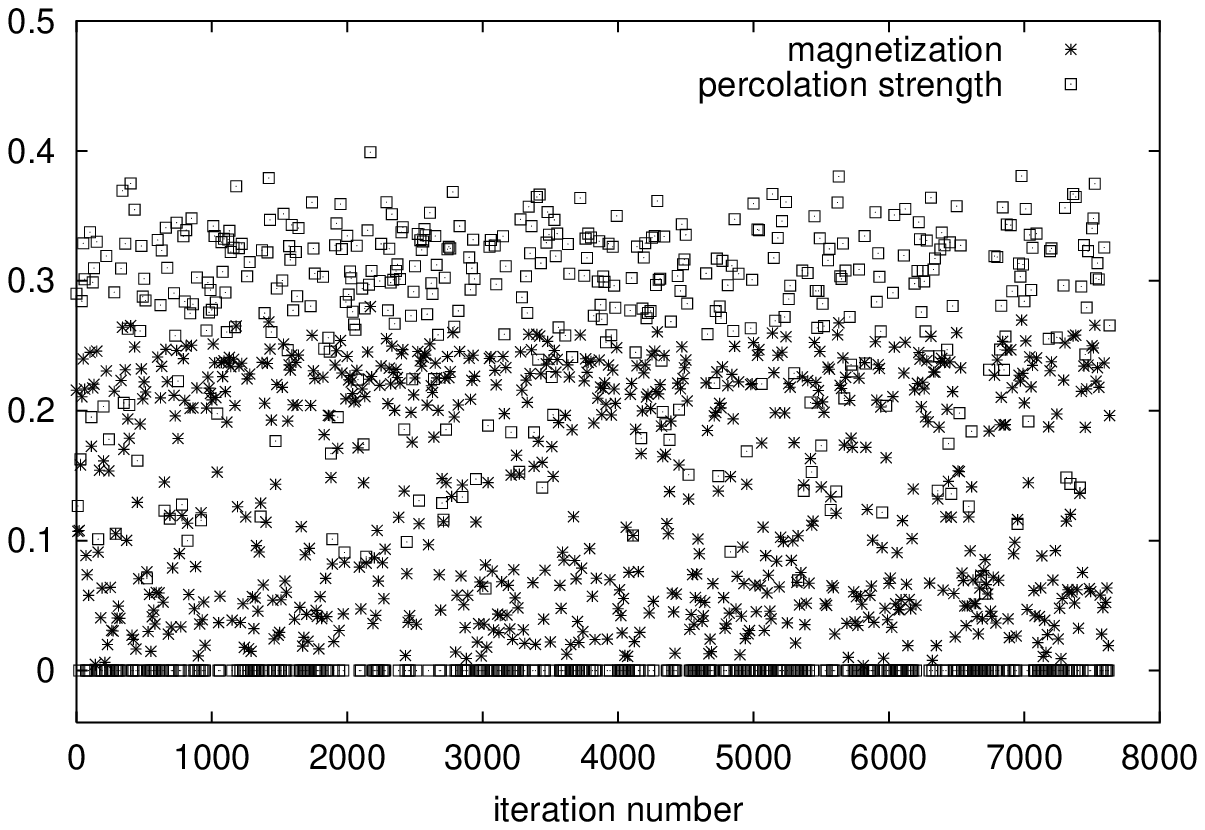,width=12.5cm}
\caption{\label{h2}{Time history at the transition point and with
field $h=0.0006$ for the magnetization $M$ and the percolation 
strength $P$ of FK clusters. The lattice size is $70^3$.}}
\end{center}
\end{figure}

In the presence of an external field, the magnetization is different
from zero at any temperature, so that it is no longer a genuine order
parameter for a thermal transition. Nevertheless, for $0 < h < h_c$,
$M$ shows discontinuous behaviour at some critical temperature $T_c(h)$,
and hence it remains interesting to compare again the behaviour of $M$
and $P$ for $h\not=0$ along the line of discontinuity $T_c(h)$. This line
ends at a critical value of the field, at which the thermal transition
becomes continuous; for the 3d 3-state Potts model, $h_c=0.000775(10)$
(see \cite{sven}). In \cite{sven}, several points of the line of first
order phase transitions of the model were determined as well. We choose
two of these points, corresponding to the values 0.0005 and 0.0006 of
the reduced field $h$, and there determine the time history of $M$ and
$P$; it is shown in Figs. \ref{h1} and \ref{h2}, respectively.

From these figures we see that there still are two phases, represented
by the two bands of the magnetization values, although now $M$ is never
zero. The step between the two bands is narrower for $h=0.0006$, as it 
should be, since we are approaching the critical value $h_c$ of the 
field, at which there would
be a continuous variation of $M$. In both cases, the percolation
strength also makes a jump from zero to a non-zero value, as we had
found for $h=0$. We notice that the number of `intermediate' $P$ values
between the two bands increases when one passes from $h=0.0005$ to
$h=0.0006$, which indicates that the transition from one geometrical
phase to the other is getting smoother: we are therefore approaching a
continuous percolation transition as well.

\bigskip

\noindent{\bf 3.3 The Case \boldmath{$h=h_c$}}

\bigskip

The line of first order phase transitions terminates with a continuous
phase transition at $T_c(h_c)$. All thermal variables vary continuously
or diverge at this critical point. In particular, the magnetization
varies continuously here, and the critical exponents at $T_c(h_c)$
put the transition into the universality class of the 3d Ising model.
The continuous thermal transition suggests that also the percolation
transition of the FK clusters becomes continuous at $h=h_c$.
Nevertheless, there are in principle three possible scenarios for the
percolation transition:

\begin{figure}[tb]
\begin{center}
\epsfig{file=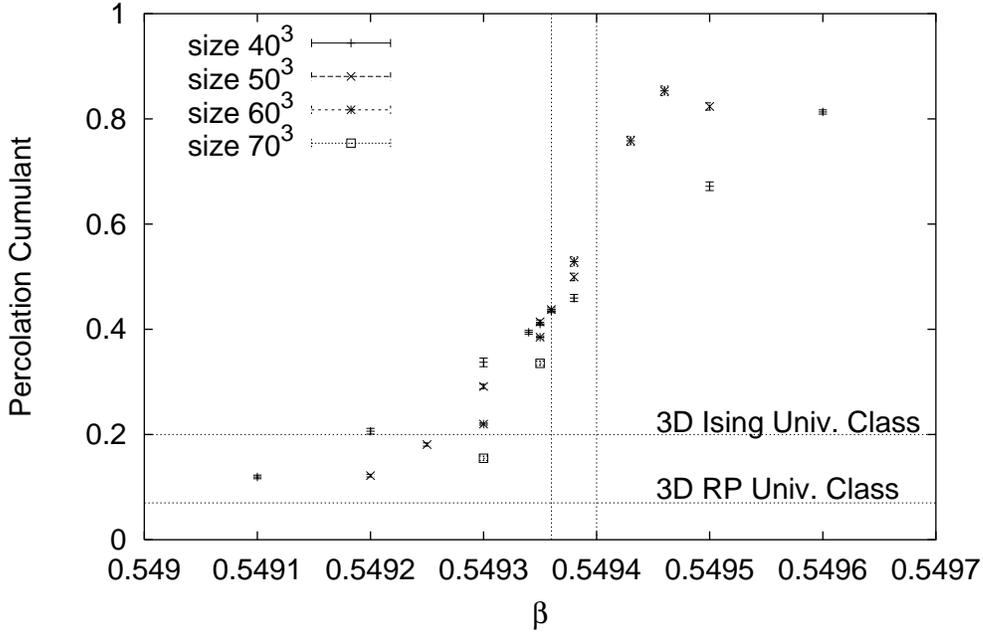,width=14cm}
\caption{\label{cum}{Percolation cumulant at the transition endpoint
$h=h_c$ for FK clusters as a function of $\beta$ for four 
lattice sizes.}}
\end{center}
\end{figure}

\begin{itemize}
\item{The FK clusters percolate at a temperature $T_p{\neq}T_c(h_c)$; in
this case, due to the finite correlation length of the thermal system at
any $T{\neq}T_c(h_c)$, the critical percolation exponents will belong to
the universality class of random percolation in three dimensions.}
\item{The configurations of FK clusters percolate at $T_c(h_c)$, but the
exponents do not coincide with the ones of the 3d Ising model, which
govern the thermal transition at this point.}
\item{The configurations of FK clusters percolate at $T_c(h_c)$, and the
exponents are the 3d Ising exponents.}
\end{itemize}

\par

We have seen so far that the threshold of the (first order) percolation
transition coincides with the (first order) thermal threshold from $h=0$
up to $h=0.0006$: that suggests that also for $h=h_c=0.000775$, the two
critical points coincide.

\par

To check if this is indeed correct, we plot in Fig. \ref{cum} the
percolation cumulant defined in Section 2 as a function of $\beta$ for
different lattice sizes. We see that within errors the lines cross at
the same point. The vertical dashed lines of the figure mark the
thermal threshold determined in \cite{sven} within one standard
deviation. The agreement between percolation and thermal critical
temperatures is very good, particularly if we take into account
that we performed our simulations for $h_c=0.000775$, even though
the value of the critical field $h_c$ also contains some uncertainty.

\par

From Fig.\ \ref{cum} we also obtain first indications of the critical
exponents of the percolation transition. The height of the crossing
point is a universal number, i.e., it identifies a universality class.
In our figure, the horizontal 
lines shown correspond to the universality classes of
the 3d Ising model and 3D random percolation (see \cite{tesi}). The
crossing point does not fall on either of the two lines, which means
that the percolation exponents of our transition coincide neither with
the 3d Ising exponents nor with the exponents of random percolation.

\par

Using standard finite size scaling techniques, we obtain the following
values for the critical percolation exponents: $\beta/\nu=0.32(3)$,
$\gamma/\nu=2.32(2)$, $\nu=0.45(3)$. From the scaling of the size of
the largest cluster at $T_c$ we find that the fractal dimension $D$ of
the percolating cluster is $D=2.66(3)$. It is easy to check the our
values satisfy the scaling relations
\be
\frac{\gamma}{\nu}+2\,\frac{\beta}{\nu}=d, \hskip2cm
\frac{\gamma}{\nu}+\frac{\beta}{\nu}=D,
\ee
within the errors we have determined; here $d$ is the space dimension
of the lattice. However, the critical indices we have found differ from the
Ising ($\beta/\nu=0.5187(14)$, $\gamma/\nu=1.963(7)$, $\nu=0.0.6294(10)$) 
and from the random percolation exponents ($\beta/\nu=0.477(2)$, 
$\gamma/\nu=2.045(10)$, $\nu=0.0.8765(17)$), as expected.
The fact that our exponents do not coincide with the random percolation
exponents is in fact a further proof that the geometrical transition
takes place exactly at the thermal threshold, because only the
presence of an infinite correlation length can shift the values of
the critical indices out of the random percolation universality class.

\bigskip

\noindent{\bf 4.\ Conclusions}

\bigskip

We have shown that FK clusters reveal interesting complementary
critical features also for spin models undergoing a first order phase
transition. In particular, we find that for the 3d 3-state Potts model,
the percolation strength jumps from zero to a non-zero
value for any value of the external field up to the endpoint.
Therefore, the line of
thermal first order phase transitions is also a line of first order
percolation transitions for the FK clusters, and the percolation
strength constitues a genuine order parameter for any
$0\,{\leq}\,h<h_c$, in contrast to the magnetization. For $h=h_c$, the
percolation transition becomes continuous, and its threshold coincides
with the thermal critical point. However, the critical indices of the
geometrical transition are not in the 3d Ising universality class.

\par

The value of $D$ we have found is bigger than the fractal dimension
of the percolating FK cluster in the 3d Ising model 
($D_{\rm Ising}=2.48(2)$);
hence the FK clusters seem too large to define correctly the thermal
critical behaviour of the model at the endpoint. This is reminiscent of
the situation encountered for the pure site clusters of the 2d Ising
model, which reproduce the right critical temperature \cite{con1} but
not the exponents \cite{sykes}. In that case the introduction of the
bond probability $p_B$ reduces the size of the clusters and restores
the correct critical behaviour. We thus conclude that also in our case
one has to define a correct bond probability $p_B(J,H,T)$ to obtain a
coincidence of thermal and percolation transitions.

\par

In this work, we have limited ourselves to the 3d 3-state Potts model, 
mainly because here the endpoint is known precisely. As mentioned 
in Section 3, much computer time is needed to obtain uncorrelated 
configurations, and hence it would have been very time-consuming to 
investigate further systems. Nevertheless, it would be of interest 
to check whether the results we have found are valid for any Potts 
model which undergoes a first order phase transition for $h=0$.

\vskip1.5cm

\noindent{\bf Acknowledgements}

\bigskip

It is a pleasure to thank F. Karsch and S. Stickan for helpful
discussions. We would gratefully acknowledge the financial support of
the TMR network ERBFMRX-CT-970122 and the DFG Forschergruppe FOR
339/1-2.

\end{document}